# Avoided level crossing at the magnetic field induced topological phase transition due to spin-orbital mixing

G. Krizman[1,2], B.A. Assaf[1], M. Orlita[3,5], T. Phuphachong[2], G. Bauer[4], G. Springholz[4], G. Bastard[2], R. Ferreira[2], L.A. de Vaulchier[2], Y. Guldner[2]

[1] Département de Physique, Ecole Normale Supérieure, PSL Research University, CNRS, 24 rue Lhomond, 75005 Paris, France
[2] Laboratoire Pierre Aigrain, Département de Physique, Ecole Normale Supérieure, PSL Research University, Sorbonne Université, CNRS, 24 rue Lhomond, 75005 Paris, France
[3] Laboratoire National des Champs Magnétiques Intenses, CNRS-UJF-UPS-INSA, 38042 Grenoble, France
[4] Institüt für Halbleiter- und Festkörperphysik, Johannes Kepler Universität, Altenbergerst. 69, 4040 Linz, Austria
[5] Institute of Physics, Charles University, Ke Karlovu 5, 12116 Praha 2, Czech Republic

Abstract: In 3D topological insulators, an effective closure of the bulk energy gap with increasing magnetic field expected at a critical point can yield a band crossing at a gapless Dirac node. Using high-field magnetooptical Landau level spectroscopy on the topological crystalline insulator $Pb_{1-x}Sn_xSe$, we demonstrate that such a gap closure does not occur, and an avoided crossing is observed as the magnetic field is swept through the critical field. We attribute this anticrossing to orbital parity and spin mixing of the N=0 levels. Concurrently, we observe no gap closure at the topological phase transition versus temperature suggesting that the anticrossing is a generic property of topological phase transitions.

The search for Dirac fermions beyond 2D [1] [2] [3] [4] has stimulated investigations of tunable topological material [5] [6] [7] [8] [9] [10] [11] [12] that possess an energy gap that can be varied from negative through zero to positive using external knobs as illustrated in Fig. 1(a). The zero-gap state at the phase boundary between the trivial and topological phase is expected to be a realization of a critical 3D Dirac state. The thermodynamic stability of this critical point is essential to our fundamental understanding of topological phase transitions [13] [14] [15] [16] [17] and our realization of Dirac fermions beyond 2D. One particularly striking feature of topological materials is the inverted behavior of their N=0 Landau levels (LL) versus magnetic field. [4] [9] [18] [19] [20] [13] In such systems, the N=0 conduction LL decreases in energy when the magnetic field increases, whereas the N=0 valence level increases (see Fig. 1(b)). This leads to an effective closure of the energy gap and a topological phase transition at a critical field $B_c$. Although previously studied in 2D for HgTe quantum wells, [13] [21] [22] [23] this transition has not been observed in 3D topological systems as most of those have large energy gaps requiring $B_c$ in excess of 100T. [24] This problem can be alleviated using IV-VI topological crystalline insulators (TCI), in which the energy gap can be tuned close to zero by choice of the proper composition. [7] [8] [9] [11] [14]

In this work, we therefore study the critical behavior of the 3D TCI $Pb_{1-x}Sn_xSe$ as a function of magnetic field in the vicinity of the critical point $B_c$ of the topological transition, which occurs at B=25T for x=0.19. [11] The mirror-like band structure [11] [19] of $Pb_{1-x}Sn_xSe$ yields linear N=0 LLs in the entire magnetic field range of interest, and can thus result in a clear cut determination of $B_c$ and the behavior around it. Using a detailed analysis of magnetooptical transitions and their oscillator strengths, we thus demonstrate the presence of an avoided-crossing of the N=0 conduction and valence Landau levels at the critical field $B_c$ as illustrated in Fig. 1(c). This anti-crossing unambiguously manifests itself via the appearance of otherwise forbidden magnetooptical transitions that clearly violate conventional selection rules. We attribute this violation to the presence of spin-orbital mixing of the N=0 LL near $B_c$.

We also show that the anticrossing is present at all temperatures in the topological regime as well as for the topological phase transition as a function of temperature. Considering the crystalline and dielectric properties of $Pb_{1-x}Sn_xSe$, we can rule out (i) bulk inversion asymmetry, (ii) electron-phonon interactions (iii) electron-electron interactions and (iv) surface effects [14] as a possible origin for the anticrossing. We thus discuss our findings in light of recent proposals on the role of alloy disorder [17] and fundamental thermodynamic effects that yield a topological phase transition without a gap closure. [25]

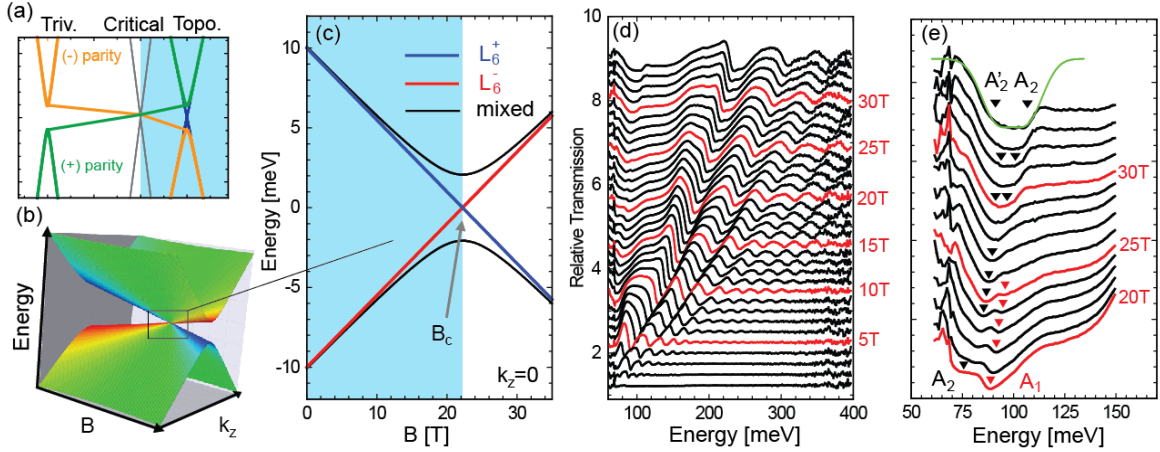

FIG 1. (a) Topological phase transition from trivial to topological with the critical gapless state shown in grey. The blue shade indicates the topological regime. (b) Evolution of the bulk N=0 Landau levels as a function of magnetic field for a topological system. The color indicates the $L_6^{\pm,\downarrow,\uparrow}$ orbital character of the N=0 levels. (c) Zoom-in on the region at $k_z$=0 showing crossing LL with pure $L_6^\pm$ character in blue and red and the anti-crossing LL with mixed $L_6^\pm$ character in black. (d) Magnetooptical transmission (T(B)/T(0)) spectra measured in $Pb_{1-x}Sn_xSe$ (x=0.19) versus magnetic field between 2T and 34T at 1.6K. Red curves mark magnetic fields that are multiple of 5T and the curves are arbitrarily shifted vertically for clarity. (e) Zoom-in on the low energy region between 20T and 34T highlighting the behavior of three particular transitions labeled $A_1$, $A_2$ and $A_2'$. A multi-peak fit shown in green is used to separate $A_2'$ and $A_2$ at high fields.

Magnetooptical Landau-level spectroscopy measurements are performed on high-quality [111]-oriented $Pb_{1-x}Sn_xSe$ epilayers grown by molecular beam epitaxy as described in our previous work. [11] [26] Two 2-μm thick samples with composition x=0.19 and x=0.14, respectively above and below the topological phase transition at 4.2K, are studied. Magnetospectroscopy is performed in transmission mode in the Faraday geometry versus magnetic field and temperature. We use a setup analogous to one used in our previous works [27] [11] with a composite Si bolometer mounted behind the sample for measurements below 4.2K at the Laboratoire National des Champs Magnétiques Intenses in Grenoble up to 34T. The probe is equipped with ZnSe windows with an energy cutoff close to 60meV. Temperature dependent measurements up to 200K are performed at Ecole Normale Supérieure up to 17T using a second setup with external detectors (far-IR bolometer or HgCdTe mid-IR detector).

Figure 1(d) shows magnetooptical spectra obtained for $Pb_{0.81}Sn_{0.19}Se$, between B=2T and B=34T at 1.6K, up to 400meV. Pronounced transmission minima originating from LL transitions are observed. A zoom-in on the low energy section of the spectra for fields between 20T and 34T is shown in Fig. 1(e). A strong transmission minimum ($A_1$) marked by the red arrow is observed below 100meV and decreases in amplitude as B approaches 25T. A second minimum ($A_2$) occurring at lower energy marked by a black arrow gains in amplitude and becomes dominant at very high fields. This transition widens above ~28T, as it splits into two transitions. A multi-peak fit allows to separate the two transitions, and pin-point the position of A$_2$ and A$_2$' as shown in Fig. 1(e). The behavior observed in Fig. 1(e) is unique to the topological sample and is *not* observed for the topologically trivial Pb$_{0.86}$Sn$_{0.14}$Se sample as shown in the supplement. [28]

Next, we will show that the three transmission minima and their changing amplitudes and energies versus magnetic field are direct evidence of the topological phase transition that occurs after an avoided crossing of the valence and conduction N=0 LLs at B$_c$ (Fig. 1(c)). To this end, we compute the LL spectrum using the k.p method introduced by Mitchell and Wallis [29] detailed in our previous work. [11] In Pb$_{1-x}$Sn$_x$Se the band extrema occur at the L-points of the Brillouin zone and the orbital basis near the band edges consists of two bands of opposite parity referred to as $L_6^{\pm}$. [29] [30] [31] The interaction between these two bands perturbed by higher order bands that are farther away from the band-edge [29] [31] [32] yields a massive Dirac-like band structure. [11] [9] [33] In the topological state, the conduction band is $L_6^+$ and valence band $L_6^-$. [34] [7] In this case, the LL energy is given by:

$$E_{N>0}^{c,\uparrow/\downarrow} = \pm\hbar\widetilde{\omega} + \sqrt{(\Delta - \hbar\widetilde{\omega}N)^2 + 2v_c^2\hbar eBN}$$

$$E_{N=0}^{c,\downarrow} = \Delta - \hbar\widetilde{\omega}$$

$$E_N^{v,\uparrow/\downarrow} = \pm\hbar\widetilde{\omega} - \sqrt{(\Delta - \hbar\widetilde{\omega}N)^2 + 2v_c^2\hbar eBN} \qquad (1)$$

$$E_{N=0}^{v,\uparrow} = -\Delta + \hbar\widetilde{\omega}$$

Here $\Delta(>0)$ is the half-energy gap, $v_c$ is the Dirac velocity related to the k.p matrix element and $\hbar\widetilde{\omega} = \hbar eB/\widetilde{m}$ where $\widetilde{m}$ includes the far-band contribution to the band-edge mass and the g-factor $\tilde{g} = |2m_0/\widetilde{m}|$. [9] N is the Landau index, $c/v$ is the conduction/valence band index and the $\downarrow/\uparrow$ index denotes the 'effective spin' introduced by Mitchell and Wallis. [29] [30] [28] [35] The LLs are plotted versus magnetic field in Fig. 2(a) up to 34T. The N=0 levels are linear in B and (anti-) cross at B$_c$ i.e. when $\hbar\widetilde{\omega} = \Delta$. Beyond B$_c$ the two levels interchange.

Note that only transitions pertaining to the levels of the oblique valleys of (111)-oriented Pb$_{1-x}$Sn$_x$Se are considered as they are known to be dominant in the optical absorption for x=0.19. The valley anisotropy is very small [31] rendering the valley splitting only visible at very high fields. Interband magnetooptical transitions between the LLs given in Fig. 2(a) verify the selection rules ΔN=±1 and conservation of effective spin, [31] [29] [28] [11] so that the transition energies are those of a massive Dirac spectrum:

$$E_N^{c,\uparrow/\downarrow} - E_{N\pm 1}^{v,\uparrow/\downarrow} = \sqrt{(\Delta - \hbar\widetilde{\omega}N)^2 + 2v_c^2\hbar eBN} + \sqrt{\left(\Delta - \hbar\widetilde{\omega}(N\pm 1)\right)^2 + 2v_c^2\hbar eB(N\pm 1)} \quad (2)$$

For N=1, we get a transition energy from N=0$^v$ to N=1$^{c,\uparrow}$ referred to as the A$_1$-transition (see Fig. 2(a)). For B<B$_c$ i.e. $\hbar\widetilde{\omega} < \Delta$, this is the dominant interband transition. As N=0$^v$ is ideally pure of effective spin and orbital character($L_6^{-,\uparrow}$), [29] the transition to N=1$^{c,\uparrow}$ is the only one allowed if the conventional [31] [32] [11] selection rules are obeyed. Similarly, the only intraband transition that is allowed is the N=0$^c$ to N=1$^{c,\downarrow}$ referred to as the A'$_1$ transition (see Fig. 2(a)). Two additional transitions, the N=0$^{v,\uparrow}$ to N=1$^{c,\downarrow}$ and the N=0$^{c,\downarrow}$ to N=1$^{c,\uparrow}$ shown as dashed arrows (A$_2$ and A'$_2$) are not allowed as they do not conserve the effective spin, if the N=0 levels do not hybridize.

The hybridization of the N=0 levels results in an anticrossing which renormalizes their energy:

$$E_{N=0}^{c/v} = \pm\sqrt{(\Delta - \hbar\widetilde{\omega})^2 + W^2} \qquad (3)$$

Here, W is the hybridization energy. The renormalized N=0 levels computed using Eq. (3) are shown in Fig. 2(a) with the respective orbital weight illustrated by a color gradient. Accordingly, the $A_1$-transition is shifted in energy:

$$E_{A1} = \sqrt{(\Delta - \hbar\widetilde{\omega})^2 + 2v_c^2\hbar eB} + \sqrt{(\Delta - \hbar\widetilde{\omega})^2 + W^2} \quad (4)$$

This hybridization also leads to spin and orbital mixing of the two levels and to the activation of the $A_2$ and $A'_2$ transitions near $B_c$ as shown in Fig. 2(a).

From the data shown in Fig. 1(d,e), we construct a fan diagram and fit the magnetooptical transitions with the calculated LLs. One free fit parameter ($W$) is needed for transitions involving the N=0 levels. Using transitions involving higher N, the energy gap and velocity are found to be $2\Delta = 20 meV$, $v_c = 4.7 \times 10^5 m/s$ (see Fig. 2(b) and supplement [28]). $\widetilde{m}$ is fixed to $0.28 m_0$, in agreement with previous studies [31] [19] and to yield $B_c \approx 24T$, the field at which $A_1$ and $A_2$ are almost equal in amplitude. The transition energies for $A_1$, $A_2$ and $A_2'$ are plotted versus magnetic field in Fig. 2(b) along with *N>1* transitions. Curve fits using theoretically calculated transitions that take into account the renormalized N=0 levels yield an excellent agreement with experiment for $W = (5 \pm 1) meV$ as shown in Fig. 2(b).

For $B<B_c$, the $A_1$-transition is strongest, as seen in Fig. 1(e). We do not observe the $A_1'$ intra-band transition in this regime, since it falls in an energy region that is not within experimental reach (below 60meV). $A_2$ emerges near 17T and gains in amplitude relative to $A_1$. $A_2'$ then splits from $A_2$ close to 28T. This splitting most likely corresponds to the point where the energy separation between $A_2$ and $A_2'$ becomes larger than the transition linewidth. For $B>B_c$, $A_2$ and $A_2'$ gain in amplitude and become dominant while $A_1$ is suppressed (Fig. 1(e)) since the N=0 LLs alter their spin and orbital character as seen in Fig. 2(a,b).

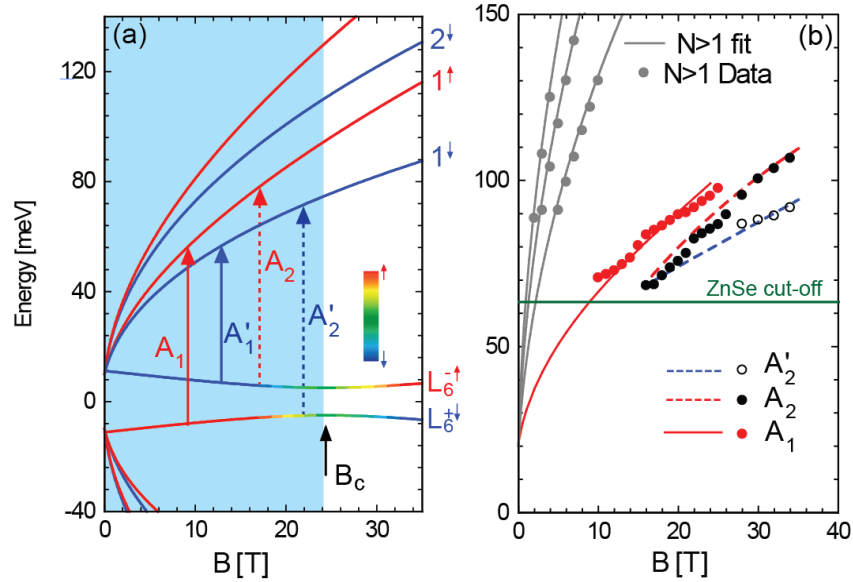

FIG 2. (a) LL energy versus magnetic field computed using Eq. 1. The effective spin 'up' ('down') levels are plotted in red ('blue'). The hybridized N=0 levels are plotted with a color gradient that illustrates their spin-orbital mixing. $B_c$ is the critical field. The $A_1$, $A_2$, $A_1'$ and $A_2'$ transitions are shown as full or broken arrows in red and blue. The blue shade indicates the topological regime. (b) Magnetooptical transition fan-chart at low energies up to B=34T. Red dots correspond to the $A_1$ absorption data points, the full black ones to $A_2$ and the empty circles to $A_2'$. The solid red curve corresponds to the calculated variation of $A_1$ and the dashed red and blue curves to that of $A_2$ and $A_2'$, respectively. The grey dots and lines are data points and curve fits using Eq. 2 for transitions not involving the N=0 levels (N>1). The green line represents the cutoff of the ZnSe Grenoble probe window.

In order to shed light on the orbital nature of the N=0 levels, the ratio of the oscillator strength of the A$_1$ and A$_2$ transitions is computed using their matrix elements at each field. This ratio is mainly imposed by the evolution of the $L_6^{+,\downarrow}$ to $L_6^{-,\uparrow}$ component for the N=0$^{c/v}$ Landau level. It is derived in the supplement [28] and given by:

$$\left(\frac{A_2}{A_1}\right) = \frac{\left(\Delta - \hbar\widetilde{\omega} - \sqrt{(\Delta - \hbar\widetilde{\omega})^2 + |W|^2}\right)^2}{|W|^2} \quad (5)$$

Eq. (5) allows us to quantify the mixing of the N=0 levels. Fig. 3(a) shows the comparison between Eq. (5) and the experimental absorption amplitude ratio A$_2$/A$_1$ extracted by fitting the spectra shown in Fig. 1(e). The agreement is remarkable, up to 25T, the field at which A$_1$ can still be reliably extracted. The consequences of this result are of major fundamental importance, as they allow *a direct experimental determination of the changing spin-orbital character of each level*, and therefore of the effective evolution of the topological state of the system with magnetic field. When the ratio is smaller than 1, the system is non-trivial meaning that the conduction N=0 is dominantly $L_6^{+,\downarrow}$ and the valence level $L_6^{-,\uparrow}$. When it exceeds 1, the character is reinverted and trivial parity is smoothly restored.

Note that we have only probed the transition that occurs for the oblique valleys of Pb$_{1-x}$Sn$_x$Se. $\widetilde{m}$, and therefore B$_c$, are is slightly valley dependent. [31] Hence, the transition for the longitudinal valley can occur at a field slightly lower than for the oblique ones. Accordingly, there has to exist an intermediate phase where the system is topologically similar to a Z$_2$-topological insulator with three inverted bands, before becoming trivial.

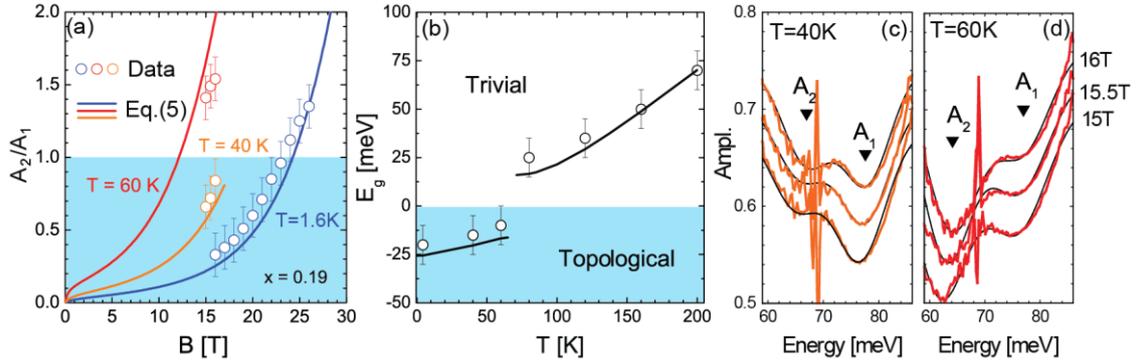

FIG 3. (a) Relative amplitude of A$_2$ to A$_1$ (empty circles) compared with the relative oscillator strength (solid lines) calculated using Eq. 5. for T=1.6K, 40K and 60K for Pb$_{0.81}$Sn$_{0.19}$Se. (b) Energy gap versus temperature for Pb$_{0.81}$Sn$_{0.19}$Se. Empty circles are data points and the solid line is calculated using Eq. 6. (c,d) Magnetooptical spectra measured between 15T and 16T for T=40K and 60K respectively. The A$_1$ and A$_2$ transitions are highlighted. A Gaussian-broaded multi-peak fit to the data is shown in black.

Measurements at higher temperature (up to 200K) demonstrate the evolution of A$_1$ and A$_2$ through the topological phase transition in Fig. 3. Topological Pb$_{1-x}$Sn$_x$Se exhibits a closure of the energy gap as temperature is increased, with a critical temperature of T≈70K for x=0.19. [14] [34] Magnetooptical measurements were performed to follow the variation of the energy gap, and track the changing amplitude of A$_2$ and A$_1$ near B$_c$ up to T=200K. Corresponding fan charts are shown in the supplement. [28] The experimental variation of the gap across the topological phase transition at T=70K is plotted in Fig. 3(b).

In the topological regime, the energy gap is equal to -15meV at 40K and -10meV at 60K (Fig. 3(b)). Low energy spectra taken at 40K and 60K are shown in Fig. 3(c,d), highlighting the evolution of A$_2$ and A$_1$ versus temperature and magnetic field. Assuming $\widetilde{m}$ does not change with temperature we expect

$B_c \approx 18T$ at 40K and $B_c \approx 12T$ at 60K. The changing amplitude of $A_2$ and $A_1$ can be observed in both cases between 15T and 16T. At 40K, B=16T remains smaller than $B_c$, therefore $A_1$ remains strongest and weakens as the field is increased as seen in Fig. 3(c). At 60K, $B_c$ drops to 12T. By 15T(>$B_c$), the system has already undertaken a partial parity re-inversion and $A_2$ becomes stronger than $A_1$ as seen in Fig. 3(d). The amplitude ratio $A_2/A_1$ for those two temperatures is shown in Fig. 3(a) and compared to the calculated oscillator strength. The parameters used for the calculation are identical to those used at 1.6K, apart from $E_g$ which is taken from Fig. 3(b). The agreement between the calculation and the data is excellent at all temperatures in Fig. 3(a), confirming the presence of spin-orbital mixing near $B_c$ for any given temperature at which the system is non-trivial.

Note that in the trivial phase (T≥70K or for $Pb_{0.86}Sn_{0.14}Se$ [28]), we do not observe the behavior attributed to the mixing since magnetooptical transitions obey conventional selection rules. [11] This further confirms our interpretation relating the emergence of forbidden transitions to spin-orbital mixing. Overall, our experimental results allow us to quantitatively extract the degree of mixing and to demonstrate a smooth topological-to-trivial transition with an anticrossing at $B_c$.

In Fig. 3(b), it is also interesting to notice the absence of gap closure for the temperature induced topological phase transition, as previously suggested by ARPES measurements on $Pb_{1-x}Sn_xSe$ single crystals. [14] Near the critical temperature $T_c$, this anticrossing $W'$ is extracted by fitting the variation of the gap versus temperature with the following empirical equation as shown in Fig. 3(b):

$$\tilde{E}_g(T) = \pm \sqrt{E_g(T)^2 + 4W'^2} \qquad (6)$$

Here, $E_g(T) = E_g(4.2K) - 32 + \sqrt{32^2 + 0.58^2 T^2}$. The $\pm$ sign is for trivial and topological respectively. We find $W' = (8 \pm 1) meV$ of the same order of magnitude as $W$, suggesting that both are caused by a single universal mechanism. Note that for 2D HgTe quantum wells the anticrossing of the N=0 LLs has been attributed either to inversion asymmetry or to electron-electron interactions. [13][19] [15] Rocksalt $Pb_{1-x}Sn_xSe$ is, however, inversion symmetric and exhibits a huge dielectric constant [36] ruling out both mechanisms. [13][19] [15] [37] Moreover, the nearly temperature independent $W$ also rules out the role of electron-phonon interactions. Previous magnetooptical work on Bi proposed a field dependent $W$ as a coupling term stemming from time-reversal-symmetry breaking. [38] [39] [40] This, however, cannot explain our anticrossing observed versus temperature. Thus, overall, we are able to retain either an extrinsic mechanism such as alloy disorder [17] or atomic vacancies or a thermodynamic first-order transition as proposed in ref. [25] as the possible origin for $W$ and $W'$. Further systematic measurements on samples with controlled level of disorder and theoretical work are required to resolve this issue.

To conclude, magnetooptical measurements at high magnetic fields evidence a topological phase transition induced by the field with an avoided crossing at the critical point. The avoided crossing arises from spin-orbital mixing that occurs when the N=0 LL converge towards each other at high field. This effect is observed concurrently with a temperature induced topological phase transition without gap closure. This suggests that the absence of gap closure versus temperature and the anticrossing near $B_c$ are due to a single universal mechanism. Elucidating this mechanism for topological phase transitions without gap closure is of fundamental importance to realize critical 3D gapless Dirac modes using tunable topological materials.

**Acknowledgments.** We acknowledge fruitful discussions with Y. Fuseya, O. Pankratov, S. Krishtopenko, and F. Teppe. This work is supported by Agence Nationale de la Recherche LabEx grant ENS-ICFP ANR-10-LABX-0010/ANR-10-IDEX-0001-02 PSL and by the Austrian Science Fund, Projects P 28185-N27 and P 29630-N27. G.K.

is also partly supported by a PSL scholarship. We acknowledge the support of LNCMI-CNRS, a member of the European Magnetic Field Laboratory (EMFL).is also partly supported by a PSL scholarship. We acknowledge the support of LNCMI-CNRS, a member of the European Magnetic Field Laboratory (EMFL).

[1]   M. Neupane, S.-Y. Xu, R. Sankar, N. Alidoust, G. Bian, C. Liu, I. Belopolski, T.-R. Chang, H.-T. Jeng, H. Lin, A. Bansil, F. Chou, and M. Z. Hasan, Nat. Commun. **5**, 3786 (2014).

[2]   J. Xiong, S. K. Kushwaha, T. Liang, J. W. Krizan, M. Hirschberger, W. Wang, R. J. Cava, and N. P. Ong, Science (80-. ). **350**, 413 (2015).

[3]   I. Belopolski, S. Y. Xu, N. Koirala, C. Liu, G. Bian, V. N. Strocov, G. Chang, M. Neupane, N. Alidoust, D. Sanchez, H. Zheng, M. Brahlek, V. Rogalev, T. Kim, N. C. Plumb, C. Chen, F. Bertran, P. Le Fèvre, A. Taleb-Ibrahimi, M. C. Asensio, M. Shi, H. Lin, M. Hoesch, S. Oh, and M. Z. Hasan, Sci. Adv. **3**, 12 (2017).

[4]   M. Konig, S. Wiedmann, C. Brune, A. Roth, H. Buhmann, L. W. Molenkamp, X.-L. Qi, and S.-C. Zhang, Science (80-. ). **318**, 766 (2007).

[5]   S.-Y. Xu, Y. Xia, L. A. Wray, S. Jia, F. Meier, J. H. Dil, J. Osterwalder, B. Slomski, A. Bansil, H. Lin, R. J. Cava, and M. Z. Hasan, Science **332**, 560 (2011).

[6]   T. Sato, K. Segawa, K. Kosaka, S. Souma, K. Nakayama, K. Eto, T. Minami, Y. Ando, and T. Takahashi, Nat. Phys. **7**, 840 (2011).

[7]   P. Dziawa, B. J. Kowalski, K. Dybko, R. Buczko,  a Szczerbakow, M. Szot, E. Łusakowska, T. Balasubramanian, B. M. Wojek, M. H. Berntsen, O. Tjernberg, and T. Story, Nat. Mater. **11**, 1023 (2012).

[8]   P. S. Mandal, G. Springholz, V. V. Volobuev, O. Caha, A. Varykhalov, E. Golias, G. Bauer, O. Rader, and J. Sánchez-Barriga, Nat. Commun. **8**, 968 (2017).

[9]   B. A. Assaf, T. Phuphachong, E. Kampert, V. V. Volobuev, P. S. Mandal, J. Sánchez-Barriga, O. Rader, G. Bauer, G. Springholz, L. A. De Vaulchier, and Y. Guldner, Phys. Rev. Lett. **119**, 106602 (2017).

[10]  L. Wu, M. Brahlek, R. Valdés Aguilar,  a. V. Stier, C. M. Morris, Y. Lubashevsky, L. S. Bilbro, N. Bansal, S. Oh, and N. P. Armitage, Nat. Phys. **9**, 410 (2013).

[11]  B. A. Assaf, T. Phuphachong, V. V Volobuev, G. Bauer, G. Springholz, L.-A. De Vaulchier, and Y. Guldner, NPJ Quantum Mater. **2**, 26 (2017).

[12]  T. H. Hsieh, H. Lin, J. Liu, W. Duan, A. Bansil, and L. Fu, Nat. Commun. **3**, 982 (2012).

[13]  M. Orlita, K. Masztalerz, C. Faugeras, M. Potemski, E. G. Novik, C. Brüne, H. Buhmann, and L. W. Molenkamp, Phys. Rev. B **83**, 115307 (2011).

[14]  B. M. Wojek, P. Dziawa, B. J. Kowalski, A. Szczerbakow,  a. M. Black-Schaffer, M. H. Berntsen, T. Balasubramanian, T. Story, and O. Tjernberg, Phys. Rev. B **90**, 161202 (2014).

[15]  A. V. Ikonnikov, S. S. Krishtopenko, O. Drachenko, M. Goiran, M. S. Zholudev, V. V. Platonov, Y. B. Kudasov, A. S. Korshunov, D. A. Maslov, I. V. Makarov, O. M. Surdin, A. V. Philippov, M. Marcinkiewicz, S. Ruffenach, F. Teppe, W. Knap, N. N. Mikhailov, S. A. Dvoretsky, and V. I. Gavrilenko, Phys. Rev. B **94**, 155421 (2016).

[16]  M. Salehi, H. Shapourian, N. Koirala, M. J. Brahlek, J. Moon, and S. Oh, Nano Lett. **16**, 5528 (2016).

[17]  W. Zhang, M. Chen, J. Dai, X. Wang, Z. Zhong, S.-W. Cheong, and W. Wu, Nano Lett. **18**, 2677 (2018).

[18]   B. A. Bernevig and T. L. Hughes, *Topological Insulator and Topological Superconductors* (Princeton University Press, 2013).

[19]   A. Calawa, J. Dimmock, T. Harman, and I. Melngailis, Phys. Rev. Lett. **23**, 7 (1969).

[20]   M. Schultz, U. Merkt, A. Sonntag, U. Rossler, R. Winkler, T. Colin, P. Helgesen, T. Skauli, and S. Lovold, Phys. Rev. B **57**, 14772 (1998).

[21]   M. Marcinkiewicz, S. Ruffenach, S. S. Krishtopenko, A. M. Kadykov, C. Consejo, D. B. But, W. Desrat, W. Knap, J. Torres, A. V. Ikonnikov, K. E. Spirin, S. V. Morozov, V. I. Gavrilenko, N. N. Mikhailov, S. A. Dvoretskii, and F. Teppe, Phys. Rev. B **96**, 35405 (2017).

[22]   A. M. Kadykov, S. S. Krishtopenko, B. Jouault, W. Desrat, W. Knap, S. Ruffenach, C. Consejo, J. Torres, S. V. Morozov, N. N. Mikhailov, S. A. Dvoretskii, and F. Teppe, Phys. Rev. Lett. **120**, 86401 (2018).

[23]   M. S. Zholudev, F. Teppe, S. V. Morozov, M. Orlita, C. Consejo, S. Ruffenach, W. Knap, V. I. Gavrilenko, S. A. Dvoretskii, and N. N. Mikhailov, JETP Lett. **100**, 790 (2015).

[24]   M. Orlita, B. a. Piot, G. Martinez, N. K. S. Kumar, C. Faugeras, M. Potemski, C. Michel, E. M. Hankiewicz, T. Brauner, Č. Drašar, S. Schreyeck, S. Grauer, K. Brunner, C. Gould, C. Brüne, and L. W. Molenkamp, Phys. Rev. Lett. **114**, 186401 (2015).

[25]   V. Juričić, D. S. L. Abergel, and a. V. Balatsky, Phys. Rev. B **95**, 161403 (2017).

[26]   G. Springholz and G. Bauer, in *Semicond. Quantum Struct. - Growth Struct. Landolt-Börnstein Vol. III/34*, edited by C. Klingshirn (Springer Verlag, Berlin, 2013), pp. 415–561.

[27]   B. A. Assaf, T. Phuphachong, V. V. Volobuev, A. Inhofer, G. Bauer, G. Springholz, L. A. De Vaulchier, and Y. Guldner, Sci. Rep. **6**, (2016).

[28]   See supplementary material for fan charts of x=0.14 at 1.6K, x=0.19 at 1.6K, 40K and 200K, and the derivation of Eq. (5).

[29]   D. L. Mitchell and R. F. Wallis, Phys. Rev. **151**, 581 (1966).

[30]   R. L. Bernick and L. Kleinman, Solid State Commun. **8**, 569 (1970).

[31]   G. Bauer, in *Narrow Gap Semicond. Phys. Appl. Proceeding Int. Summer Sch.*, edited by W. Zawadzki (Springer Berlin Heidelberg, Berlin, Heidelberg, 1980), pp. 427–446.

[32]   H. Burkhard, G. Bauer, and W. Zawadzki, Phys. Rev. B **19**, 5149 (1979).

[33]   T. Phuphachong, B. A. Assaf, V. V. Volobuev, G. Bauer, G. Springholz, L.-A. de Vaulchier, and Y. Guldner, Crystals **7**, (2017).

[34]   A. J. Strauss, Phys. Rev. **157**, 608 (1967).

[35]   G. Bauer, H. Pascher, and W. Zawadzki, Semicond. Sci. Technol. **7**, 703 (1992).

[36]   H. Preier, Appl. Phys. **20**, 189 (1979).

[37]   A. A. Abrikosov and S. D. Beneslavskii, J. Low Temp. Phys. **5**, 141 (1971).

[38]   G. A. Baraff, Phys. Rev. **137**, A842 (1965).

[39]   M. P. Vecchi, J. R. Pereira, and M. S. Dresselhaus, Phys. Rev. B **14**, 298 (1976).

[40]   Z. Zhu, B. Fauqué, K. Behnia, and Y. Fuseya, **30**, (2018).